# Autonomous Vehicle Networks for More Reliable Truck Tracking in Challenged High Mountain Roads, Tunnels and Bridges Environments

Junhao Chen[1], Milena Radenkovic[2]*

[1] The University of Nottingham; psxjc12@nottingham.ac.uk
[2] The University of Nottingham; milena.radenkovic@nottingham.ac.uk
* Correspondence: milena.radenkovic@nottingham.ac.uk

**Abstract:** The popularity of online shopping has challenged the existing express tracking. How to provide customers with reliable and stable express tracking has become one of the important issues that express companies need to solve now. The current stage of courier tracking is not ideal in challenging environments such as mountain roads, tunnels and city centres. Therefore, the project aims to overcome the challenging environment and achieve stable express tracking, and proposes the Ya'an scenario and conducted multiple experiments. We show that opportunistic DTN-aware protocols are feasible solution for trucks to maintain stable communication in challenging environments, and nodes maintain extremely high message delivery rates and average delays that can maintain communication.

## 1. Introduction

With the continuous upsurge of online shopping, express logistics industry has introduced new challenges for both vehicle transport and networks communities [1]. Especially in China, according to China e-Commerce Research Centre (CECRC), from 2012 to 2016, the growth rate of China's online retail transaction volume was more than 35% every year, and even reached 50% in 2014 [2]. In 2016, the number of parcels delivered by China's express delivery reached 31.28 billion and maintained an annual growth rate of more than 50% [3]. One of the questions major express service companies are committed to optimizing [4], is how to provide customers with real-time and accurate location information of trucks and packages at a low cost.

The current solutions typically aim to get the locations of the trucks which are carrying the packages via Global Position System (GPS). While this approach may provide good real-time and accurate information [5]. Based on this feature, GPS is even used in car anti-theft and vehicle speed detection [6].

Emerging research [7] has shown that GPS signals are often blocked by high-rise buildings which leads to the vehicle losing its position when it travels in challenged areas such central areas of cities (which are often congested), as well as tunnels or mountain roads (which may be isolated). In addition to this, communications may be further disrupted in case when accidents happen. When tracking large fleets, the cost of maintaining communications increases as the number of connections to the central server increases [8]. At the same time, the size of the data packet also affects the probability of data transmission errors. The larger the data packet, the higher the probability of errors [9].

This paper investigates the hypothesis that Delay Tolerant and opportunistic Networks can be an effective means for package tracking in such challenged complex and unstable environments when trucks are moving.

In order to gain better understanding of the impact of failures of GPS signals on the road conditions and services, we propose to consider Ya'an city in China, which is surrounded by mountain roads and tunnels. We show that complex mountain roads and long tunnels cause GPS to fail for long periods of time [7]. A recent city risk report [10] shows that the central and western parts of Ya'an city are at high risk of mudslides which may further disrupt the power and communications if the city is hit by a mudslide.

This work will design and emulate the Ya'an realistic traffic and communication patterns in ONE simulator and conduct a range of experiments to study multi-dimensional protocol performance analysis for multiple protocols across multiple metrics. We assume a delivery truck (as the source node), is sending location coordinates to other neighboring nodes (other vehicles), which help the source node transfer coordinate information to the target node. And target node can be diverse. It could be a store in the town tracking the delivery of fresh fruits and vegetables. It could also be a police station, tracking post-disaster relief. And the target node of this research is set at the base stations in the center of the town. They will collect the coordinate information of the delivery truck and upload it to the database by Internet. Customers and express companies can track the delivery of the truck at any time through the networked devices.

There have been some studies deploying DTN in different cities for experiments. The research [11] uses DTN to track electric vehicles in San Francisco. By monitoring the number and remaining power of electric vehicles in the area, inferring their charging needs, and dynamically adjusting the energy supply of charging piles in different areas to make charging more efficient. The University of Massachusetts research team [12] used 35 public transport buses equipped with computers and various wireless devices, as well as several outdoor access points deployed in the city canter, to build an experimental site with a communication coverage of 150 square miles called DieselNet. And for the Mobile Opportunistic Disconnection Tolerant Networks and Systems (MODiToNeS) [13] was focus on developing a distributed architecture that supports real-time multi-layer and multi-dimensional communication, so that DTN can be better integrated with mobile social and transportation systems. The fault and disconnection aware smart sensing (FDASS) framework [14] aims to reduce the load on the network, by detecting and avoiding connections to faulty nodes in the network and directing traffic to nodes that are more reliable and less congested. While the aforementioned studies are very innovative and have a huge contribution to the development of DTN, none of them consider the challenging conditions as we did for Ya'an. Therefore, our proposed work in this paper considers a more challenging novel scenario and provides novel results and discussions.

Our paper describes the rich set of experiments in Ya'an dynamic challenged environment with varying communication and mobility patterns and discuss different routing protocols performances over different metrics. Our results show which opportunistic DTN protocols are most suitable for the challenging Ya'an scenario and discuss how this relates to other emerging studies in cave and after disaster challenging environments. We show that opportunistic DTNs can maintain stable data transmission in complex and challenged vehicle movement and communication conditions and that this allows the technology to be used to track high-value targets, such as secure-car (check word) and VIP vehicles.

## 2. Literature Review

*2.1. Background on Delay Tolerant Networks (DTNs, opportunistic networks and routing protocols)*

The Delay Tolerance Networks (DTNs) have an overlay layer between transport layer and application layer and follow store-carry-forward mechanism for messages dissemination[8] which is generally suitable for communication in the face of potentially high delays and error rates and low bandwidth [15].

As an important part of DTN protocol, bundle protocol provides the functions of custody transfer, proactive and reactive bundle fragmentation, and late binding, which is the basis for the successful implementation of DTN [16]. In the bundle protocol, the bundle is used as the basic unit of data transmission, and the convergence layer adapter is used to match the different protocols of Bluetooth and WIFI, so that DTN can solve the problems existing in heterogeneous network interconnection.

In this paper we study opportunistic DTNs (rather than scheduled DTNs or static DTNs) in which nodes are mobile, communication is contact-based [18] and routing algorithms are based on flooding and probability [19] which is most appropriate for this type of scenario.

*2.2. DTN routing protocols*

Routing protocol is the essential element of DTN to improve the delivery rate and reduce the delivery delay [20].

*2.2.1. Epidemic routing protocol*

As in the research article [21], the source node of the Epidemic protocol will send the same number of copies of messages to all nodes in its communication range. The node that gets the copy of the message then becomes the distributor of the copy of the message, distributing the message with same number of copy to other nodes within the communication range [22]. This means that there is no limit to the number of copies in a scenario for a single message. As nodes continue to forward messages, the number of copies in the scenario grows exponentially.

The biggest potential problem with an Epidemic protocol is that there is no limit to the number of copies, which can overload buffers for nodes in a scenario, leading to an information jam. In this project, this protocol mainly plays the role of experimental control group.

*2.2.2. Spray and Wait routing protocol*

Compared to the Epidemic, the SprayAndWait routing protocol limits the number of copies of messages [23]. It works in the same way as the Epidemic, with both forwarding schemes based on replication, and SprayAndWait routing protocol is arguably an improved version of the Epidemic [24]. The SprayAndWait routing protocol has two ways of assigning copies of messages, Vanilla and Binary.

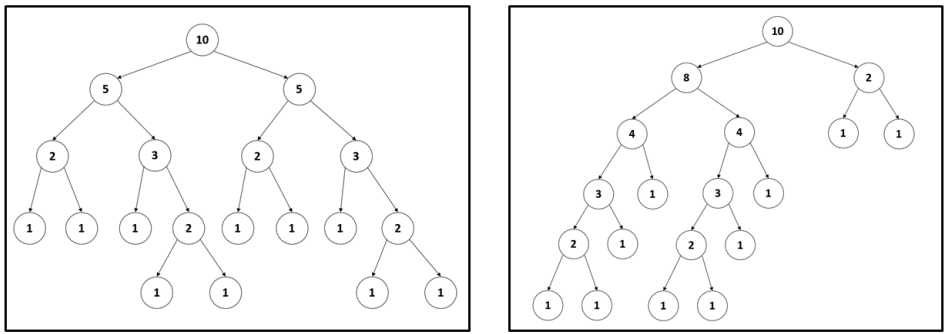

*Figure 1: Vanilla (left) & Binary (right) SprayAndWait*

As shown in the Fig 1, the node starts with 10 copies of the message, and by distributing a specified number of copies of the message to the nodes it encounters in two different ways, the node ends up with only one copy of the message. When the number of copies of the node is greater than 1, the node is in spraying stage. When the number of copies of a node is equal to 1, the node enters waiting phase, reserving the copy until it encounters the destination node.

However, Spray and Wait also has its limitations. When the movement range of nodes in the area is limited, the node carrying the copy may never reach the destination node, such as the school area [26]. On the other hand, the initial number of copies (L) of the source node is also an important parameter affecting Spray and Wait performance [27]. The value of 'L' is related to the number of nodes in the scenario, and the appropriate value needs to be determined through simulation experiments.

*2.2.3. MaxProp routing protocol*

The MaxProp routing agreement prioritises nodes based on the successful delivery rate of node messages and packet drop rate. The research shows that [28], the MaxProp routing protocol performs better than other protocols when the node's schedule trajectory is known. MaxProp prioritizes packets based on the number of hops. The lower number of hops, the higher the priority, ensuring fast propagation across the network. When packets exceed the hop count threshold, priority is reassessed based on the probability that the two peers meet, using incremental averaging. Through information exchange between nodes, confirm and delete copy of the message has been delivered [29], saving buffer space and improving delivery efficiency.

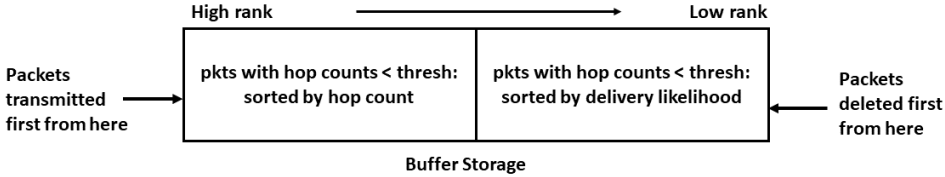

*Figure 2: Strategy of MaxProp routing*

It will be interesting to see how the protocol performs in the experiment, given that the delivery trucks have a rough route before departure. But there are challenges, such as the possibility that trucks could be rerouted in the middle of an emergency and whether communications would be affected. These are the items that need to be considered in the scenario layout of the project, and add relevant events in the simulation experiment to observe the influence on the experimental results.

*2.3. Opportunistic Network Environment (ONE) Simulator*

ONE simulator is a discrete event simulator based on Agent [30]. Compared to other simulators, it integrates moving models, DTN routing, and a visual graphical interface. This makes the ONE high rate of scalable, allowing users to customize scenarios as needed. In addition, ONE can provide a large number of results reports and analysis models for the project to analyse experimental data [31]. Considering that the project will simulate specific scenarios and conduct comparative experiments on different protocols, the features of ONE can meet the requirements of the project.

In ONE, a simulation environment consists of node movement simulation, routing simulation, visual interface and message reporting module. The movement models can be generated by the integrated movement module or manually imported through the external interface. The message events generated by the simulation can be exported to the report module for further analysis.

The node movement behaviour in the scenario is realized through movement Models. Project will mainly use MapBasedMovement to plan the movement of nodes in the scenario. MapBasedMovement reads and caches map data [31]. In other words, it controls the mobility path of nodes base on map data. Different mobility paths can be set for different types of nodes. The production of project scenario map data and arrangement of node paths will be introduced below.

### 2.4. Emerging Challenging DTN Case Study

#### 2.4.1. Cave Security Monitoring

Authors in [32] propose that DTN could solve communication problems in caves . The experiments were carried out in the Postojna Cave in Slovenia. Several automatic weather stations are installed in the cave as source nodes. When the train carries passengers through the cave, the train will act as a carrying node to collect the cave meteorological data of the measurement stations along the way, and transmit them to the target node train station to establish data communication.

The results of the experiment are gratifying, the meteorological data from the depths of the cave can be successfully transmitted to the end point by train. Due to the complex structure of the cave, it is very difficult and expensive to collect meteorological data in the cave by manual or other traditional methods. Solving the communication in the cave through DTN greatly reduces the manpower and material resources. The only fly in the ointment is that the automatic measurement equipment placed in the cave during the experiment has battery life issues and requires regular maintenance. But there is no doubt that DTN is a viable solution to cave communication.

#### 2.4.2. Post-disaster Communications

A survey has shown that, due to the special network architecture of DTN, its performance is particularly bright in the scene after natural disasters [33]. Because natural disasters are often accompanied by damage to conventional communication facilities and power supply facilities, local and large-scale communication loss often occurs in disaster-stricken areas. This will cause the rescue team to be unable to accurately understand the disaster situation in the disaster area and dispatch effective rescue. How to maintain the communication in the disaster area has always been a problem that scholars have been struggling to solve.

The study [33] proposes to use DTN to maintain communications in disaster areas. The network architecture will consist of four layers. The first layer consists of rescue workers (RW) who collect and pass data to the second layer of throw-box (TB). TB is a temporary master control station that distributes collected data to passing data mules (DM), such as cars, trains, ferries and so on. Finally, the DM of the third layer transmits the data to the main control station (MCS) of the fourth layer, and the MCS connects the data with the outside world. As shown below.

The study [33] simulated post-disaster scenarios and how the DTN would work under these conditions with ONE Simulation. The experimental results prove that DTN can maintain communication in the case of regional traditional communication interruption after disaster. And experiments were carried out on different protocols, and a new one was proposed. The protocol decides whether to deliver a copy of a message based on factors such as the node's adaptability to message delivery, the likelihood of packet drops, and buffer space. In this way, a more suitable node is selected as the next hop. This has a good reference value for the project to simulate sudden accidents.

Another study [34] proposes the Trust-Aware Communications in Disaster (TACID) framework for emergencies without substantial recovery after a disaster. It is designed to receive more useful messages (distress signals), during the prime time of rescue, helping rescue work more efficient and accurate.

TACID framework is multi-layered and multi-dimensional. The physical layer and network layer in the framework are composed of heterogeneous user groups in the disaster area. The geo-spatial-temporal physical layer only allows user groups with functional equipment to carry out sporadic Jump communication, on top of this is the social self-network layer. A node's social self-network layer records its past encounters and performance, including similarity, signal strength, and centrality. By analysing these data, infer the future reliability of the node.

The TACID framework has good adaptability, predictability and scalability, and can perform trust-aware heterogeneous communication in the presence of hostile nodes. But at the moment it has not addressed requests for emergencies that require urgent handling, but they undoubtedly present a solid solution to post-disaster communications.

#### 2.4.3. Contact tracing of infectious diseases

Traditional contact tracing of infectious diseases requires a lot of manpower for investigation and statistics through self-report of traced contacts. At the same time, the reliability of self-reporting is very limited due to people's limited memory, confidentiality and other personal privacy reasons. A French research team proposes to use DTN to trace contacts of infectious diseases [35].

In this experiment, 36 experimenters carried wireless modules and acted as mobile nodes. When two mobile nodes move into each other's communication range, they will record their current time and synchronize to the most recently visited server node. The experimental team set up 11 fixed nodes in the public places of the school, 3 connected to the Internet as server nodes, and transmitted the collected data of mobile nodes to the database. In order to confirm whether the mobile node leaves the communication range, the fixed node in the scene will broadcast a message every 4 seconds, and the mobile node will broadcast a message every 30 seconds. If the node in the current communication fails to receive 4 consecutive messages, the node will mark as out of range. At the same time, the distance between nodes can be inferred by the received signal intensity. The graph below shows the average exposure time collected by the database at different times of the day

After interviews and statistics, the database has a high degree of agreement with the data reported by the experimenter. It proves that DTN can provide us with highly reliable contact tracing, which saves a lot of manpower and material resources compared to traditional traceability. The layout of DTN experiment scenario in this study has important reference value for our project. At the same time, it also provides new ideas for the application of DTN in real world.

*2.4.4. Aircraft cluster network*

When the fleet is conducting combat missions, the traditional wireless network will be disconnected frequently due to the high-speed movement of the aircraft, the complex and changeable geographical environment and the signal interference of the enemy. A study [36] proposes to introduce DTN into the cluster network. The special network architecture of DTN allows communication between different subnets. And the drones, which act as messenger nodes, can move on demand, passing information directly to the next subnet node.

But they still face many challenges. The existing DTN routing is difficult to adapt to the high real-time performance required by the high-speed movement of the aircraft, and cannot efficiently complete data transmission. On the other hand, the existing DTN security mechanism is far from the traditional terminal-based security mechanism, and is far from meeting the existing security standards. In addition, the DTN involves multiple information exchanges and forwarding on the way to transmit messages, during which there are extremely high security risks. This is what DTN still needs to improve in the cluster network.

**3. Ya'an Truck Package Tracking**

*3.1. Ya'an city for package tracking via DTN vehicles*

This paper studies how feasible and effective using opportunistic DTN vehicle communication compared to using traditional GPS location when tracking of the trucks carrying express goods while moving on mountain roads and tunnels in Ya'an city scenario. Ya'an city is highly challenging because it is surrounded by mountains, where winding mountain roads and tunnels are the main traffic arteries. We design realistic truck mobility, connectivity, and communication patterns in Ya'an city scenario which is driven by the real-world map of Ya'an and realistic loss of GPS signal which results in variable and potentially long disconnection times.

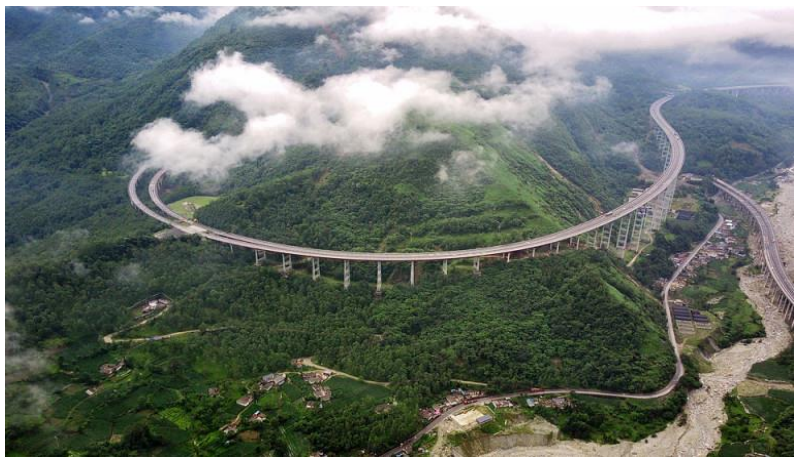

*Figure 3 :* Ya'an—Xichang Expressway [37]

Figure 3 shows the Ya'an section of Xichang Expressway, which is called 'Road on cloud' because it is built around mountains and erected much higher than the road surface. However, conventional communications equipment does not work because of the difficulty of deploying communication infrastructure in mountainous areas [38].

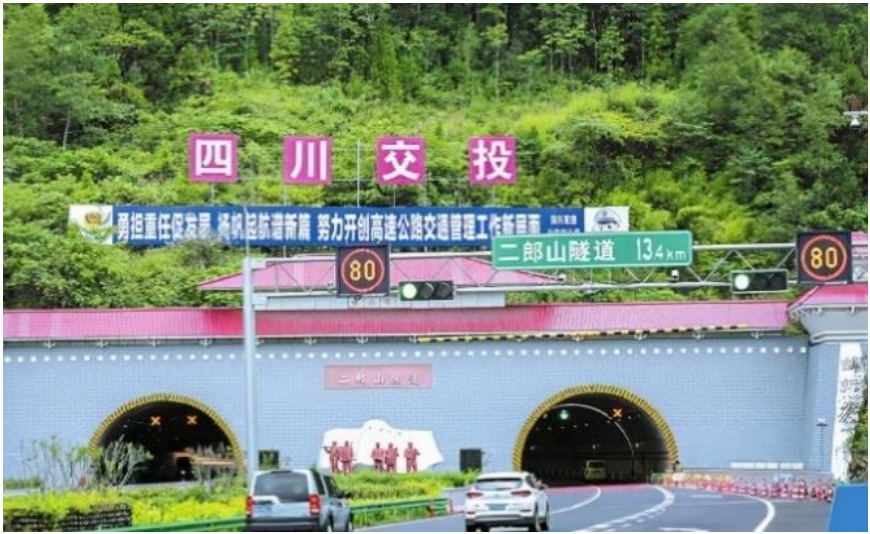

*Figure 4:* Erlangshan Tunnel [39]

Figure 4 shows one of the longest tunnels in Ya'an city, total length of 13.4 KM, called Erlangshan Tunnel. This means that when a vehicle enters the tunnel, it can't be located by GPS for more than a quarter of an hour. And there are many more tunnels in Ya'an, so it is important to find a way, instead of GPS, to locate vehicles in the tunnels.

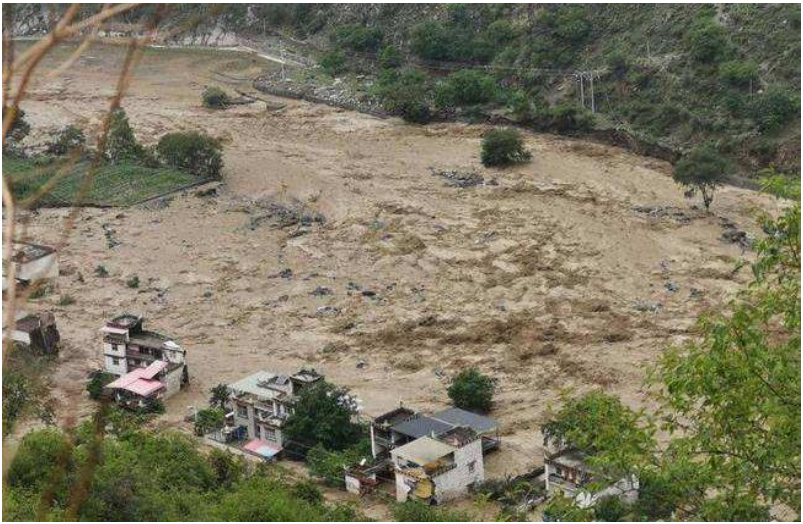

*Figure 5:* Debris flow in Ya'an, Sep 2021 [40]

At the same time, Ya'an is also a high incidence of debris flow zone [10]. Once a mudslide occurs, it will cause widespread communication disruption. The intermittent connection and tolerate of DTN can help in the emergency rescue natural disasters and communication report on follow-up rescue work [41].

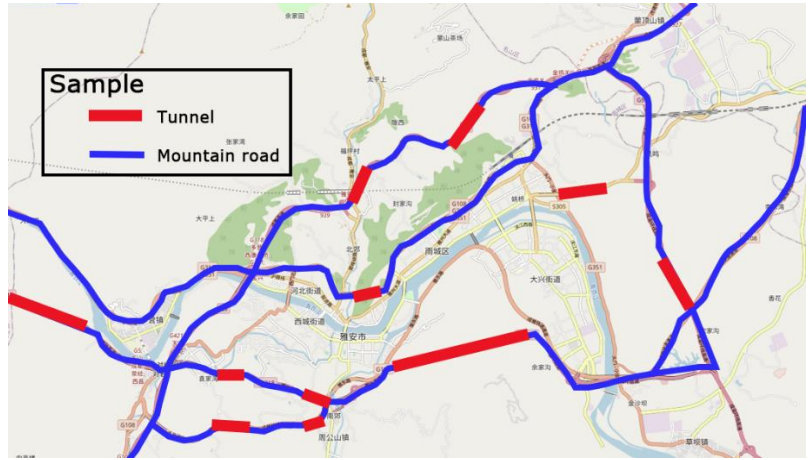

*Figure 6: Mountain Roads & Tunnels around Ya'an*

Figure 6 marks the mountain roads and tunnels surrounding Ya'an City. Due to the lack of infrastructure in these sections, conventional communication methods will not work stably. This means that when the trucks travel on these road sections, the positioning will be unstable, or even the positioning will be lost.

In order to conduct realistic experiments for Ya'an city in ONE simulator, we obtained the topological map of the city via OpenStreetMap. OpenStreetMap is a resource collective that provides user-generated road maps, allowing users to export real-world road data [42]. The map files exported through OpenStreetMap are in '.osm 'format, while ONE simulator can only recognize the map files in '.wkt ' format [31]. Well-known Text (WKT) is a Text markup language that allows the exchange of geographic data in textual form. 'osm2wkt' is a Java-based plug-in that converts '. osm ' files to '. WKT ' files [43]. Missing landmarks can also be checked and fixed. Because the ONE simulator does not allow breakpoints in map files. But, due to the optimization of the algorithm, if the map file size is large, it will take a long time to convert.

Each set of nodes in the scenario has an independent path of mobility. We use OpenJUMP software [44] to draw the corresponding path according to the actual mobility path of the node, based on the main map. Specific parameter settings of nodes in experimental scenarios will be explained below.

*3.2. Ya'an Heterogeneous DTN trucks, pedestrians, cars, cargo ships and base stations setting*

We design and build our Ya 'an challenging scenario to investigate package tracking in Ya 'an city and ran for 8 hours. We performed more experiments for longer and shorter number of hours (4h,8h,12h, 15h) were done but we show this without loss of generality. The location of delivery trucks is updated every minute, and size of each copy of the message is between 10 Kilobytes to 100 Kilobyte. Since we decided to use the 801.11p standard, the size of the message copy will be much larger than just sending the truck coordinates. The 801.11p standard is a standard Vehicular delay-tolerant networks (VDTNs), which can send bundled messages [45]. The advantage of the 801.11p standard is that in addition to recording the coordinate position of the node, it also records the speed and acceleration of the node's movement and message events [46]. Message events can warn other vehicles when an accident occurs. Given that the project is tracking autonomous vehicles, more real-time information about the vehicles helps us ensure their safety and stability.

The interface type of node is WIFI, communication range is 100 meters, transmit speed is 7.5 Megabytes per second. The total number of nodes in the Ya'an scenario is 76. The number of nodes will be changed according to the requirements of the experiment. The specific configuration of nodes will be explained below. The routing protocols involved in the experiment are described as follows: Epidemic, Spray and Wait, MaxProp. The routing protocol is tested independently. The scenario contains only one DTN routing protocol. The following figure shows the configuration of the scenario.

| Parameter | Value |
| --- | --- |
| Map data | Ya' an.wkt(Ya' an, China) |
| Simulation time | 28800s (8h) |
| Message per minute | 1 |
| Message size | 10kb to 100kb |
| Interface type | WIFI |
| Communication range | 100m |
| Transmit speed | 7.5mb/s |
| Routing protocols | Epidemic, Spray and Wait, MaxProp |

*Table 1:* Ya'an truck tracking parameters

Our experiments are driven by real world mobility, connectivity and communication patterns and have five different node categories (trucks, pedestrians, cars, boats, static nodes) each having their own mobility paths, speeds, connectivity patterns.

The first node category includes delivery trucks that serve as the source nodes. They send their location message to the base station in the city. Their range of movement is outside the town, around the mountain roads and tunnels, do not enter the centre of town (as shown in Fig 7). These nodes are moving with speeds ranging from 6 to 35 miles per hour. Since the speed limit of trucks in Ya'an City is 45 miles per hour on highways and tunnels [47]. However, the roads in Ya'an city are more winding and complicated with many uphill sections, trucks often do not reach the speed limit. So, project did not choose the highest limit speed as the upper limit of the source nodes.

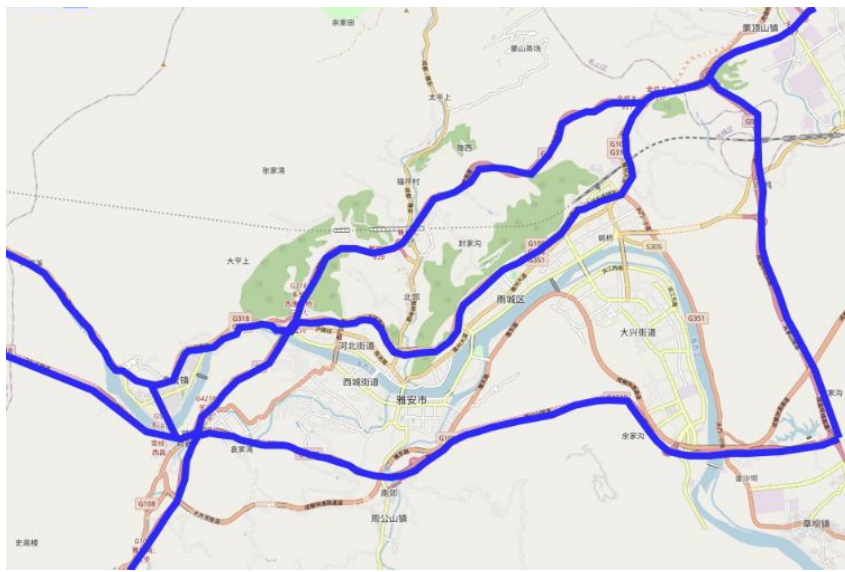

*Figure 7:* Mobility path of tracks in Ya'an

The second node category includes pedestrians with communication devices in towns who act as DTN data ferrying/carrying nodes to help with mobile ad hoc routing (as in Fig 8). Pedestrians are typically moving within the town, without leaving the town, walking at a slow pace, about 3 to 5 miles per hour. According to statistics [48], the average walking speed of an adult is about 3 miles per hour. Considering that the population density of Ya'an is low [49], the streets are rarely congested, and the walking speed of pedestrians will be slightly faster than that of the city, so the project has obtained the above speed range.

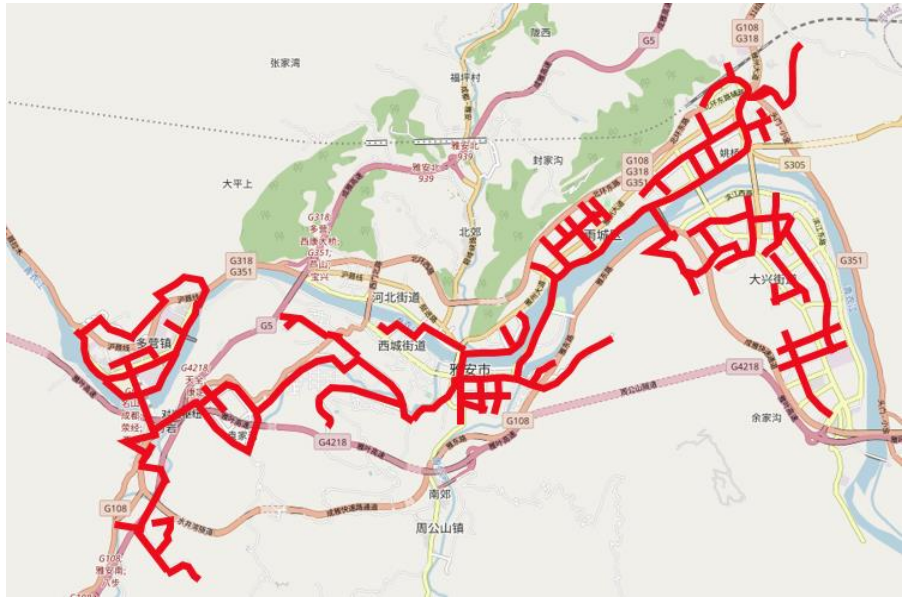

*Figure 8: Mobility path of pedestrians in Ya'an*

Third node category involves cars whose mobility pattern is similar to the trucks' mobility, but they also can move to and from the centre of the city. Cars typically move faster than trucks with speeds in the range of 8 to 45 miles per hour and act as data carrying nodes helping with the mobile ad hoc DTN routing in Ya'an scenario. The speed limit for small vehicles on highways and tunnels in Ya'an is 60 mph, and the urban speed limit is 18 mph [47]. At the same time, some roads in the urban area of Ya'an lack maintenance [50], and the roads are uneven, making it difficult for cars to maintain the maximum speed limit.

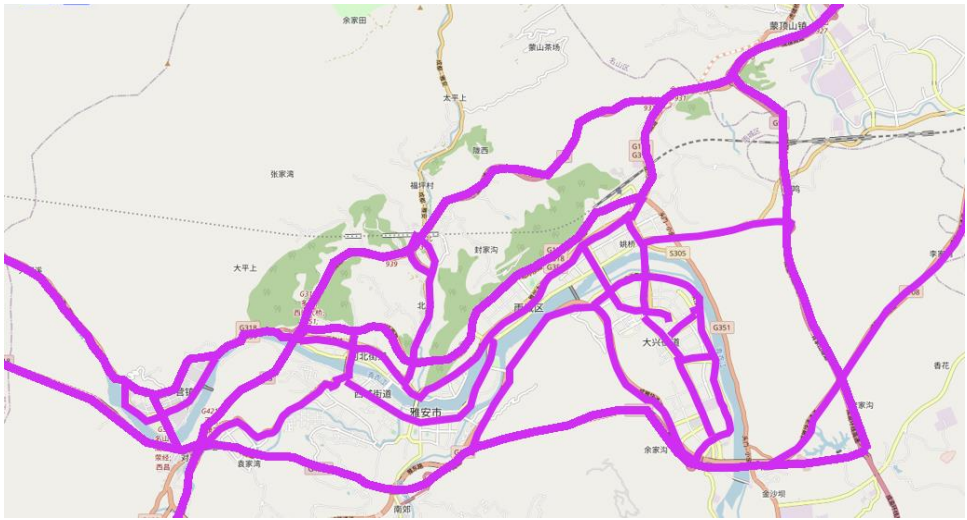

*Figure 9: Mobility path of vehicles in Ya'an*

The fourth type of node involves the frequent upstream and downstream cargo ships on Yuxi River in the north of the city [51]. The cargo ships speeds typically range between 12 and 33 mile per hours, and they act as message carrying nodes to transmit copies of the messages to the cars traveling along the highways close to the river. According to statistics [52], the average speed of freighters is about 15 miles per hour, but this value will fluctuate greatly in actual sailing due to the influence of river flow speed and heading.

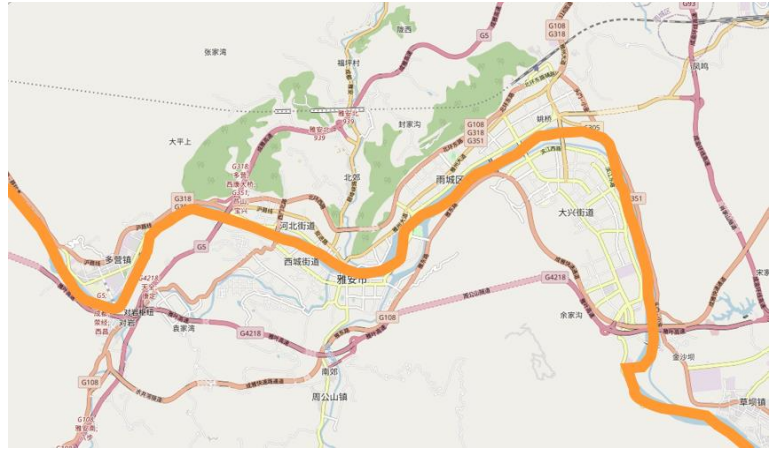

*Figure 10: Mobility path of Cargo Ships along Yuxi River in Ya'an*

Fifth category of nodes include two base stations serving as the target nodes which are located in the west and east of the town at coordinates (9600, 1000) and (5180, 10393) respectively. The interface of the base stations has a wider reception range (600m) and faster transmission speed (15mb/s) than other nodes compared to the other node categories (trucks, cars, pedestrians, ships).

The following table shows the node parameters in the Ya'an scenario.

| Name of node | Number of nodes (default) | Movement Model | Movement Speed | Comment | Group | Automatically assigned ID |
|---|---|---|---|---|---|---|
| Base Stations | 2 | StationaryMovement | 0 miles per hour | Target node | 1,2 | 0,1 |
| Trucks | 20 | ShortestPathMapBasedMovement with 'truck_paths.wkt' | 6 to 35 miles per hour | Source node | 3 | 2 to 21 |
| Pedestrians | 20 | ShortestPathMapBasedMovement with 'pedestrian_paths.wkt' | 3 to 5 miles per hour | Carrying node | 4 | 22 to 41 |
| Freighter | 5 | ShortestPathMapBasedMovement with 'freighter_paths.wkt' | 12 to 33 miles per hour | Carrying node | 5 | 42 to 46 |
| Vehicles | 30 | ShortestPathMapBasedMovement with 'vehicle_paths.wkt' | 8 to 45 miles per hour | Carrying node | 6 | 47 to 76 |

*Table 2: Node parameters for Ya'an scenario 1*

### 3.3. Design of Hybrid VDTN DTN Ya'an experiments

We perform multi-dimensional protocol performance analysis for 3 benchmark and state of the art protocols Epidemic (benchmark), Spray and Wait and MaxProp (state of the art protocols) to investigate which ones are more successful in Ya'an scenario.

We begin analysing the performance of the 3 protocols (Epidemic, Spray and Wait, MaxProp) in the Ya'an scenario. This will give us a better understanding of how feasible and successful each protocol is in the first Ya'an scenario.

For varying number of carrying nodes, we observe adaptability of the protocol for these node fluctuations in the second Ya'an scenario, which allows ad hoc DTN routing to be feasible for both peak and trough periods.

Then we investigate the performance of changing the initial number of message copies (L), and to find the 'best solution' for the Ya'an scenario.

### 3.4. Evaluation criteria

We analyse performance of multiple protocols (Epidemic, Spray and Wait, MaxProp) for a range of metrics (delivery ration, delays, overheads rate) in multiple Yann scenario(for varying number of nodes, kinds of nodes, routing protocol parameters)

$$Delivery\_rate = \frac{Delivered}{Created}$$

Delivery rate Indicates the ratio of the total number of successfully delivered messages to the total number of generated messages. Delivered in the formula refers to the total number of messages successfully Delivered to the target node, and Created refers to the total number of messages generated in the network. The project statistics are the delivery rate of trucks to base stations in the Ya'an scenario. If the delivery rate is too low, it means that there are more disconnection and congestion in the network.

$$\text{Overhead\_ratio} = \frac{\text{Relayed} - \text{Delivered}}{\text{Delivered}}$$

Overhead rate represents the ratio of the total number of additional messages forwarded in order to successfully deliver messages to the total number of successfully delivered messages in the network. We calculate the overhead rate of message delivery from the trucks to the base stations in Ya'an scenario. The smaller this parameter is, the higher the message forwarding efficiency is. Relayed in the formula represents the total number of forwarding times of messages forwarded by intermediate nodes, and Delivered represents the total number of messages successfully delivered to the destination node in the network. Excessive overhead rate represents a lot of waste of resources, which needs to be improved by replacing routing protocols or optimizing algorithms.

$$\text{latency\_avg} = \frac{\sum_{i=0}^{n-1}(t_{id} - t_{is})}{n}$$

Average delay indicates the average time it takes for all successfully delivered messages to reach the destination node from the source node. Project counts the average delay between the trucks and base stations. In the formula, $t_{id}$ is the moment when message $i$ is received by the target node, $t_{is}$ is the moment when message $i$ is generated by the source node, and n is the total number of messages successfully delivered to the target node in the network. Lower latency means that the network has a faster response speed. The latency is affected by many factors in the scenario and can be used as a reference for the overall quality of the network.

## 4. Data analysis

*4.1. Three routing protocol in Ya'an scenario*

|  | Epidemic | SprayAndWait | MaxProp |
| --- | --- | --- | --- |
| **Delivery rate** | 0.9646 | 0.9625 | 0.9646 |
| **Overhead rate** | 73.1857 | 18.3658 | 61.2419 |
| **Average delay** | 743.4756 | 843.7186 | 743.4698 |

*Table 3: Data result of Three routing protocol in Ya'an scenario*

From the experimental results in the above Table 3, the delivery rates of the three routing protocols are basically satisfactory, all greater than 0.96. But on this basis, the overhead rate of SprayAndWait is much smaller than the other two routing protocols. The main reason for this result is that it limits the number of message replicas. But this also causes the average delay of SprayAndWait to be slightly higher than the other two routing protocols. On the other hand, MaxProp is optimized through the algorithm. On the premise of ensuring the average delay, the overhead rate is reduced by about 8% compared with Epidemic.

*4.2. Heterogeneous in Ya'an scenario*

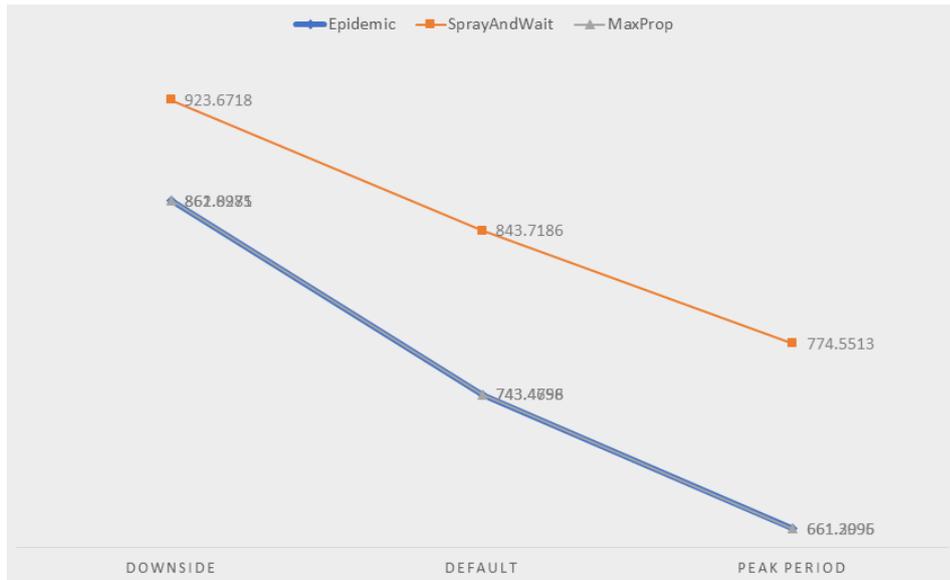

*Figure 11:* *Average Delay of Heterogeneous Ya'an scenario*

We investigate latency for varying number of nodes, as Fig 11 shows, the average latency of the three routing protocols, which is affected by the number of nodes in the scenario. An interesting phenomenon is that the average latency of Epidemic and MaxProp is almost the same in the case of three node numbers, which appears to be almost completely overlapping in the graph. From the graph we can confirm that the fewer nodes in the scenario, the higher the average latency and vice versa. This is mainly because in an opportunistic network, the more nodes there are, the greater the chance of encounters between nodes, and the faster the transfer of message copies. By observing the trend in the chart, we can see that SprayAndWait has a smoother curve than the other two routing protocols. This means that it is relatively less sensitive to changes in the number of nodes and more adaptable to changes in the scenario. But the overall average latency is higher.

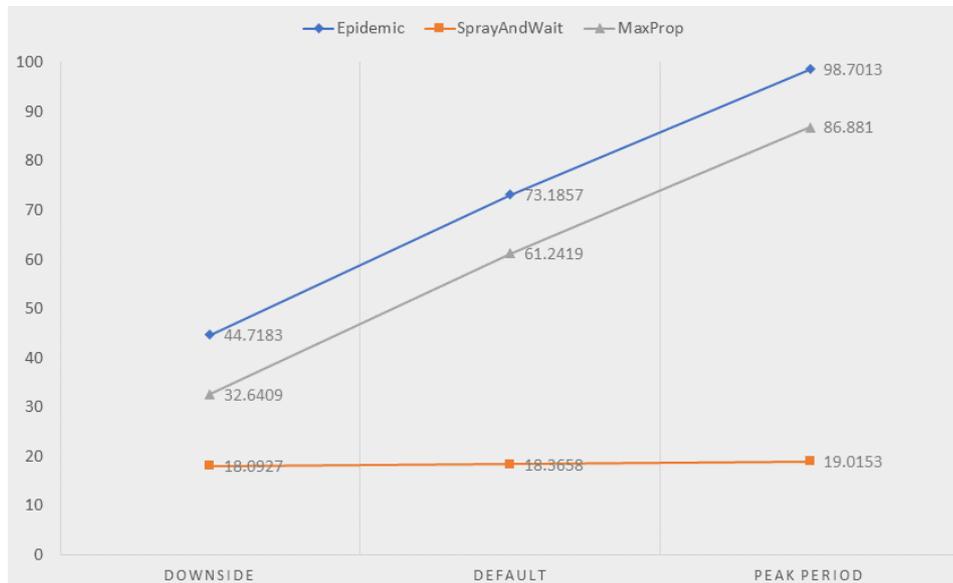

*Figure 12:* *Overhead rate of Heterogeneous Ya'an scenario*

As can be seen from Fig 12, when the number of nodes in the scenario increases, the overhead rates of Epidemic and MaxProp increase significantly. This value even exceeds 85% when the node is at its peak, which means that the copy of the message sent by the source node passes through more carrying nodes before reaching the destination node. This makes the same work, consuming more buffer resources. To optimize this problem, it is necessary to optimize the algorithm of message copy forwarding and formulate a more sensible forwarding strategy.

In contrast, SprayAndWait, because the total number of copies of the same message in the scenario is limited, when the number of copies carried by the node has one copy, the forwarding of the copy will be stopped. So even though the number of nodes in the scenario is changing, as long as the number of nodes is not less than the threshold, the overhead rate will not change significantly

|  | Epidemic | SprayAndWait | MaxProp |
|---|---|---|---|
| **Downside** | 0.9688 | 0.9667 | 0.9688 |
| **Default** | 0.9646 | 0.9625 | 0.9646 |
| **Peak Period** | 0.9625 | 0.9542 | 0.9625 |

*Table 4: Delivery rate of Heterogeneous Ya'an scenario*

From the data in the table, we can see that the message delivery rate of this experiment is satisfactory, and all the results are greater than 0.95. As the number of nodes increases, there is a very slight drop in the message delivery rate. This is mainly because the number of nodes increases, and the frequency of nodes forwarding message copies increases, resulting in nodes discarding message copies due to the full load of the buffer, which in turn reduces the message delivery rate.

*4.3. Investigate the 'L' value suitable for the Ya'an scenario*

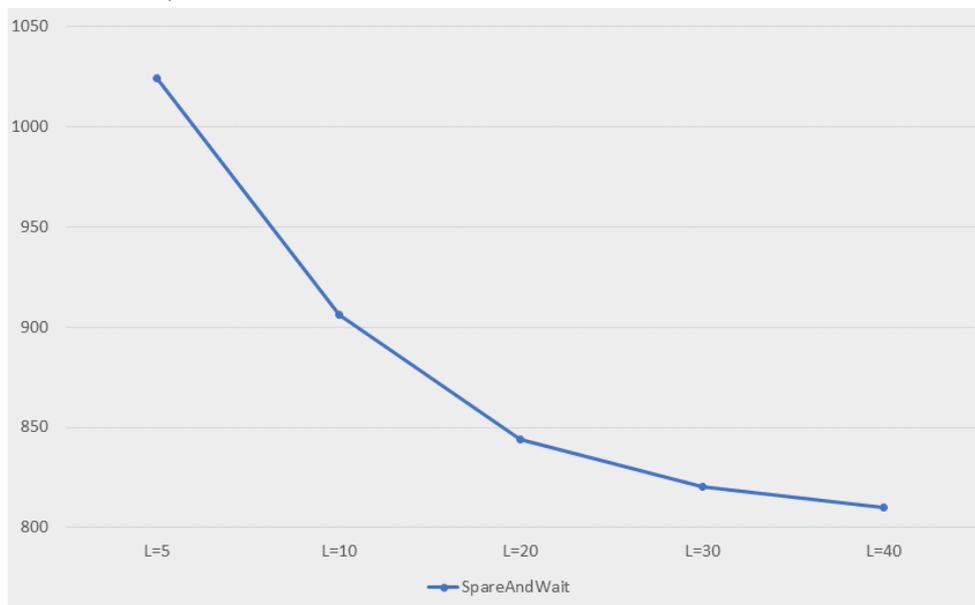

*Figure 13: Average Delay of different 'L' value in Ya'an scenario*

As Fig 13, with the value of 'L' increases, the average latency gets lower and lower. But we can clearly see that the effect of reducing the average delay by increasing the 'L' value is less and less obvious. At the same time, as the number of message copies increases, so does the risk of network congestion.

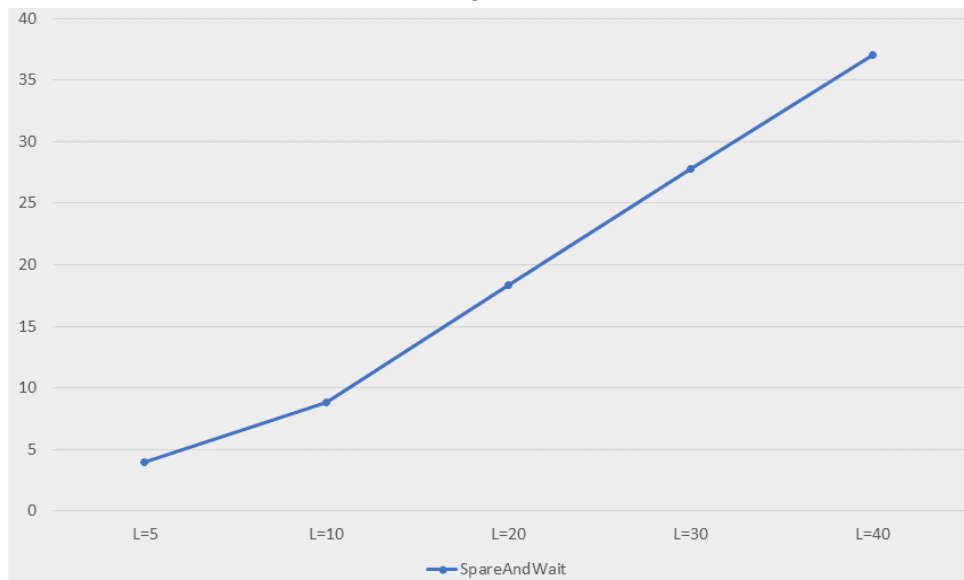

*Figure 14: Overhead rate of different 'L' value in Ya'an scenario*

As Fig 14 shown, with the increase of the 'L' value, the overhead rate shows a steady trend of increase. The larger the 'L' value, the greater the burden on the buffering of nodes in the scenario. It is not difficult to imagine that if the value of 'L' keeps rising, then its overhead rate will approach that of the Epidemic routing protocol.

Overall consideration, for the Ya'an scenario, the 'L' value of 20 is a choice that considers both the average delay and the overhead rate.

## 5. Conclusion

In this paper, we investigate a challenging Ya'an scenario based on the real Ya'an city, which restores the characteristics of Ya'an city surrounded by winding mountain roads and tunnels in the real world. Based on this Ya'an scenario, the project conducted multiple experiments around how to achieve stable and reliable packages tracking in a challenging environment.

The multiple experimental results show that the Spay and Wait routing protocol has stronger adaptability to our Ya'an scenario. When the delivery rate is almost the same, the Spay and Wait routing protocol sacrifices about 10% of the average delay compared to the other two routing protocols (Epidemic, MaxProp), but has a much lower overhead rate than the other two routing protocols. And it is not sensitive to changes in the number of nodes in the scenario, which means that it can work more stably in complex and changeable environments.

## 6. Future work

In the future work, we plan to investigate further optimization of our algorithm of the routing protocol [53] [54] based on the Ya'an scenario, so that they can work more efficiently in the Ya'an scenario, and further reduce the average delay between trucks and base stations. In addition to this, we consider adding accidental events into the Ya'an scenario to investigate the impact of emergencies on communication in the scenario. We also plan to deploy this work on realistic autonomous vehicles and drones [55] to investigate more realistic and convincing data.